\documentstyle[12pt]{article}
\def\beqar {\begin{eqnarray}}
\def\eeqar {\end{eqnarray}}
\def\beq {\begin{equation}}
\def\eeq {\end{equation}}

\begin{document}

\begin{titlepage}
\null\vspace{-62pt}

\pagestyle{empty}
\begin{center}
\rightline{CTP-3323}
\rightline{CCNY-HEP-4/02}

\vspace{1.0truein} {\Large\bf Remarks on the exotic central extension of
the}\\
\vskip .1in
{\Large\bf planar Galilei group}\\

\vspace{.5in} R. Jackiw$^a$ and V. P. Nair$^b$\\
\vskip .1in {\it $^a$ Center for Theoretical Physics\\
Massachusetts Institute of Technology\\
Cambridge, MA 02139-4307\\
\vskip .05in
$^b$ Physics Department\\
City College of the CUNY\\ New York, NY 10031}\\
\vskip .05in {\rm E-mail: jackiw@lns.mit.edu\\
vpn@sci.ccny.cuny.edu}\\
\vspace{1in}

\centerline{\large\bf Abstract}
\end{center}
Some issues in relating the central extensions of the planar Galilei
group to parameters in the corresponding relativistic theory are
discussed. 

\end{titlepage}

\pagestyle{plain}
\setcounter{page}{2}
\baselineskip =18pt
The fact that the Galilei group in (2+1) dimensions admits a
two-parameter central
extension has been known for a long time \cite{LLB}. The first of these
extensions is identified as the mass. Two years ago, we pointed out that
the second, more exotic, extension is related to the spin of the particle
\cite{JN}. This has recently been the subject of some discussion. In
\cite{hagen}, it is argued on the basis of a nonrelativistic model that
one can have spin with no extension of the algebra. The paper by Duval
and Horvathy presents an analysis of the nonrelativistic limit and the
extensions obtained by using a regularization \cite{duval}.
While we agree with the analysis of \cite{duval}, we feel
that the following remarks offer further
clarification.

The Cartan-Poincar\'e form for single particle Galilean dynamics is
given by
\beq
\omega = dp^i dx^i + {\kappa \over 2m^2} \epsilon^{ij} dp^i dp^j -
{p^i dp^i dt \over m}\label{1}
\eeq
where $m$ and $\kappa$ are the extension parameters. This form and many
of the related questions have been extensively studied \cite{all}.
The symplectic form for a free relativistic particle can be written as
\cite{JN2}
\beq
\Omega = - dp_a dx^a + {S\over 2} {\epsilon^{abc} p_a dp_b dp_c \over
(p^2)^{3\over 2} }\label{2}
\eeq
Our basic observation was that a nonrelativistic expansion of the latter
using $p_0 \approx mc + (p^ip^i /2mc)$ will lead to (\ref{1}) with
$\kappa = S/c^2$. 

Note that one may think of the nonrelativistic theory
in two
ways,
either as $p^ip^i \ll m^2 c^2$ in which case it is a nonrelativistic
approximation, or as $c\rightarrow \infty$ in which case it is a
nonrelativistic limit.  We shall consider the first way,
the approximation
$p^ip^i \ll m^2 c^2$. 
Recall that the second extension may also be characterized by the 
nonzero commutator of the Galilei boost generators $K_i$.
With $p^ip^i\ll m^2 c^2$, the
Poisson bracket of the Lorentz boost generators $J_i$ associated
with (\ref{2}) becomes
\beq
[K_i ,K_j] \approx \epsilon_{ij} (S/c^2)\label{3}
\eeq
where $J^i \approx c \epsilon^{ij} K_j$ and the right hand side arises
from the nonrelativistic approximation to $J_0$. Thus the second extension
$\kappa$ of the Galilei group is $S/c^2$.
Numerically, the appropriate dimensionless quantity
$Suv/c^2$, $u,v$ being typical velocities, may be small
in the nonrelativistic approximation. Nevertheless, $S/c^2$ remains as
an extension parameter characterizing the algebra of the Galilei boost
generators;
mathematically,
we do obtain the extended Galilei group.
If one takes $c\rightarrow \infty$, then we need
$S\rightarrow
\infty$ to get finite $\kappa$. One may feel uncomfortable with
this, but it is also mathematically correct.

We now look at the the Levy-L\'eblond equations for a spin-half particle
given by \cite{LLB2}
\beqar
i{\partial \phi \over \partial t} +p_- \chi &=& 0\nonumber\\
p_+ \phi + 2m \chi &=&0\label{4}
\eeqar
where $p_{\pm} = p_1 \pm i p_2$. In the free case, this model is
equivalent to the Lagrange density
\beq
{\cal L}=\phi^* i{\partial \phi \over \partial t} ~-~ \phi^* {p_- p_+
\over 2m}\phi \label{5}
\eeq
By assigining a new transformation law to the field $\phi$ as
$\phi\rightarrow \phi' = \exp( i {\theta \over 2}+ i\alpha \theta ) 
\phi$ under rotations, the angular momentum is seen to be shifted as $L_0
\rightarrow L_0 + \alpha \int \phi^* \phi$. Since the number operator
commutes with the Hamiltonian and various other operators of interest,
the algebra is unchanged.
Thus, in discussing nonrelativistic models, it is
important to keep in mind that, at the
level of the Galilei group, there is nothing to fix the zero of
angular
momentum. One can always add a constant without changing the algebra. 
The ambiguity of identification of spin
can never be resolved within
the Galilei symmetry. So the whole idea of relating spin to
the second extension does not work if one is limited to
the Galilean group. But this is not our point, rather we note that
there is
a definition of spin at the relativistic level and this can be
interpreted in terms of the second extension of the Galilean group.

The situation is
similar with the case of the mass. At the level of the Galilei
group, the mass is a free parameter. It is defined by
\beq
H= {p^2 \over 2m}\label{6}
\eeq
(or by the commutation rule $[K_i ,P_j]$ which is related to the
above.) There is also an additive constant we can put in
$H$, writing
\beq
H= Mc^2 +{p^2 \over 2m}\label{7}
\eeq
At the Galilean level,
there is no reason why $M$ and $m$ should be the
same; the rest mass (or rest energy)
and the inertia are independent
parameters. Relativistically, there is the 
definition of mass from the Poincar\'e group as the value of $p^\mu
p_\mu$. If we want the Hamiltonian $H$ to be the limit of this, then
$M=m$. Thus relativistic considerations give an interpretation to
the inertial mass $m$ (the first extension of the Galilei group)
as the rest mass $M$. What we are doing is to
link similarly the relativistic definition of spin to the second
extension. (In \cite{duval}, this freedom in the zero of energy and spin
is used to `renormalize' the parameters in the Galilean limit, 
explicitly carried out
by modification of the symplectic forms by including some `trivial'
extensions of the Poincar\'e group.)
 
The similarity between mass as rest energy and the spin can be seen
even
more clearly by starting from a Galilean theory which has a second
extension. As noted in \cite{brihaye, hagen}, we can achieve this
trivially by shifting the unextended Galilei boost generators
\beq
K_i \rightarrow {\tilde K}_i = K_i+{\kappa \over 2m} \epsilon_{ij} p_j
\label{8}
\eeq
This ${\tilde K}_i$ shows the second extension in its Poisson bracket
with itself. However, this modification
changes the transformation law of
$x_i$. We now have
\beq
[v\cdot {\tilde K} , x_i] = v_i t - {\kappa \over 2m} v_j \epsilon_{ji}
\label{9}
\eeq
We no longer have the expected $x\rightarrow x + vt$. Define
\beq
q_i = x_i + {\kappa \over 2m^2} \epsilon_{ji} p_j
\label{10}
\eeq
It is easily checked that $[v\cdot {\tilde K} , q_i] = v_i t$. Thus
$q_i$ is the correctly transforming coordinate, but now
$[q_i, q_j]\neq 0$. The angular momentum $L_0$
must be defined by
\beqar
[L_0 , q_i ] &=& \epsilon_{ij} q_j \nonumber\\
{}[L_0 , p_i ] &=& \epsilon_{ij} p_j 
\label{11}
\eeqar
This is easily solved to obtain
\beq
L_0 = \alpha +\epsilon_{ij} q_i p_j - {\kappa \over 2m^2} ~p^2
\label{12}
\eeq
where there is the freedom of an additive constant $\alpha$.
Note the similarity of this equation, with the two parameters
$\alpha$ and $\kappa$, to equation
(\ref{7}) for the Hamiltonian,
with the parameters $M$ and $m$. Equations (\ref{8})-(\ref{12}) define
a Galilean theory with the second extension $\kappa$, and so far, the
parameters
$\alpha$ and
$\kappa$ are not related. Now we can ask the question: can we choose the
parameters
$\alpha$ and $\kappa$
(just as we chose $M$ in terms of $m$) such that
equation (\ref{12}) is the low momentum limit
of the relativistic expression? The answer is that it can be done
if $\alpha = \kappa c^2$. But $\alpha$ is the spin from the relativistic
point of view and so we have the result of \cite{JN}. In this way of
interpreting the result, we are not taking $c\rightarrow \infty$, but
rather asking whether the {\it a priori} unrelated parameters of the
nonrealtivistic theory with the extensions can be related if viewed as
the low momentum limit of a relativistic theory.

This work was upported in part by a DOE grant number
DE-FC-02-94-ER40818, an NSF grant
number PHY-0070883 and by a PSC-CUNY grant.


\begin{thebibliography}{99}

\bibitem{LLB} J.-M. L\'evy-Leblond, in E. Loebl, {\it Group Theory and 
Applications},
Academic Press, New York (1972); D.R. Grigore, J. Math. Phys.,
{\bf 37} (1996) 460; S.K. Bose, Commun. Math. Phys. {bf 169} (1995) 385. 

\bibitem{JN} R. Jackiw and V.P. Nair, Phys. Lett. {\bf B480} (2000) 237.

\bibitem{hagen} C.R. Hagen, Phys. Lett. {\bf B539} (2002) 168.

\bibitem{duval} C. Duval and P.A. Horvathy, hep-th/0209166.

\bibitem{all} J. Lukierski, P.C. Stichel and W.J. Zakrzewski, Ann. Phys.
(NY) {\bf 260} (1997) 224; 
C. Duval and P.A. Horvathy, Phys. Lett. {\bf B479} (2000) 284.  

\bibitem{JN2} R. Jackiw and V.P. Nair, Phys. Rev. {\bf D43} (1991) 1933;
see also B. Skagerstam and A. Stern, Int. J. Mod. Phys. {\bf A5} (1990)
1575; M.S. Plyushchay, Phys. Lett. {\bf B248} (1990) 107;
Nucl. Phys. {\bf
B362} (1991), 54; A. Ballesteros, M. Gadella and M. del Olmo, J. Math.
Phys. {\bf 33} (1992) 3379; S. Ghosh, Phys. Lett. {\bf B338} (1994) 235.

\bibitem{LLB2} J.-M. L\'evy-Leblond, Comm. Math. Phys. {\bf 6} (1967)
286.

\bibitem{brihaye} Y. Brihaye, C. Gonera, S. Giller and P. Kosinski,
hep-th/9503046.

\end{thebibliography}
\end{document}